\documentclass[onecolumn]{aa} % for a paper on 1 column  
\usepackage{hyperref}
\hypersetup{
    colorlinks=true,
    linkcolor=red,
    filecolor=magenta,      
    urlcolor=magenta,
    citecolor=blue,
}
\usepackage{multirow}
\usepackage{txfonts}
\begin{document}

   \title{On the source sizes of type II radio bursts with LOFAR}

  % \subtitle{}

   \author{A. Kumari
          \inst{1, 2, 3}
   \and
    D.~E.~Morosan \inst{2,4} 
    \and
        V. Mugundhan \inst{5}
    \and
    P. Zhang \inst{6,7} 
    \and
  J.~Magdalenić \inst{8} 
    \and
    P. Zucca \inst{9}
               \and
    E.~K.~J.~Kilpua \inst{2} 
    \and
    F. Daei \inst{2}
          }

   \institute{
              Udaipur Solar Observatory, Physical Research Laboratory, Dewali, Badi Road, Udaipur-313 001, Rajasthan, India\\
               \email{anshu@prl.res.in}
   \and
   Department of Physics, University of Helsinki, P.O. Box 64, FI-00014, Helsinki, Finland 
         \and
        Solar Physics Lab, NASA Goddard Space Flight Center, Greenbelt, MD 20771, USA 
        \and
             Department of Physics and Astronomy, University of Turku, 20014, Turku, Finland
        \and
            %Onsala Space Observatory, Department of Space, Earth and Environment, Chalmers University of Technology, Onsala 439-92, Sweden
            Dept. of Space Planetary Astronomical Sciences and Engineering (SPASE), IITK, Kanpur 208016, India
        \and
        Center for Solar-Terrestrial Research, New Jersey Institute of Technology, Newark, NJ 07102, USA
        \and
        Cooperative Programs for the Advancement of Earth System Science, University Corporation for Atmospheric Research, Boulder, CO, USA
        \and
        Department of Mathematics,  Faculty of Science, KU Leuven, Oude Markt 13, 3000 Leuven, Belgium
        \and
         ASTRON, The Netherlands Institute for Radio Astronomy, Oude Hoogeveensedijk 4, 7991 PD Dwingeloo, The Netherlands
}

 %  \date{Received September 15, 1996; accepted March 16, 1997}

% \abstract{}{}{}{}{} 
% 5 {} token are mandatory
 
  \abstract
   %context heading (optional)
   {Solar radio bursts can provide important insights into the underlying physical mechanisms that drive the small and large-scale eruptions on the Sun. Since metric radio observations can give us direct observational access to the inner and middle corona, they are often used as an important tool to monitor and understand the coronal dynamics.}
   {While the sizes of the radio sources that can be observed in the solar corona is essential for understanding the nature of density turbulence within the solar corona and its subsequent influence on the angular broadening observed in radio source measurements, the smallest radio sources associated with solar radio bursts have so far been limited by observational techniques and the radio instrument's baselines.}
   %The sizes of radio sources are important in understanding the density turbulence in the solar corona, which can directly affect the intrinsic radio source sizes. This article aims to estimate the observable source sizes of type II radio bursts with the Low Frequency Array (LOFAR) observations.}
   {We selected three type II bursts that were observed with the LOFAR core and remote stations in the Solar Cycle 24. We estimated the sizes and shapes (ellipticity) of the radio sources between $20-200$ MHz using a two-dimensional Gaussian approximation.}
   {Our analysis shows that the smallest radio source size for type II bursts in the solar corona which can be observed in the solar atmosphere at low frequencies is $1.5^\prime \pm 0.5^\prime$ at 150 MHz. However, even though the observations were taken with remote baselines (with a maximum distance of $\sim 85~km$), the effective baselines were much shorter ($\sim 35~km$) likely due to snapshot imaging of the Sun.}
   {Our results show that the radio source sizes are less affected by scattering than suggested in previous studies. %We found source sizes at least $\sim$2 times smaller than previously reported values using smaller baselines suggesting that the scattering limit has not yet been reached at least in the case of type II radio bursts. 
Our measurements indicate a smaller source sizes at frequencies below 95 MHz compared to previous reports, though some overlap exists with measurements at higher frequencies, using smaller baselines. }


   \maketitle
%
%-------------------------------------------------------------------

\section{Introduction}
\label{sec:section1}

The coronal magnetic field plays an essential role in the formation, evolution, and dynamics of the small and large-scale structures in the solar corona \citep{Dulk1970, mclean1985solar}. %Gopalswamy2006}. 
These structures may lead to gigantic explosions in the solar atmosphere in the form of large-scale eruptions, such as coronal mass ejections \citep[CMEs;][and the references therein]{hundhausen1999coronal}
%webb2012coronal}
that may severely impact near-Earth space. CMEs can reach Earth within several hours to days, and depending on the orientation of its internal magnetic field; they can interact with the Earth's magnetosphere causing severe geomagnetic storms \citep{gosling1990coronal, Kilpua2014}. Moreover, the shocks generated by CMEs can accelerate the energetic particles leading to highly energetic solar radiation storms \citep[e.g.,][]{cane1984type,Morosan2024}.
%reames1995solar, 
These eruptions are often accompanied by radio emissions called solar radio bursts \citep[SRBs][]{kundu1984observations}. %Vourlidas2020}. 
Radio emission can be generated either by electrons accelerated during the eruption or thermal electrons interacting with the background plasma \citep[e.g.,][]{Morosan2019b}. These emissions can be observed with ground and space-based observatories. Radio observations are one of the most common approaches to diagnosing non-thermal electrons in the solar atmosphere \citep{2000ApJ...528L..49M, Carley2020, 2020ApJ...893..115C, Majumdar2021}. 

One direct signature of CME-driven shocks at radio wavelengths is known as metric type II radio bursts \citep{Roberts1959, Smerd1975, kumari2017b, morosan2023}. Type II bursts originate from plasma waves converted into radio waves at the local plasma frequency and/or its harmonics. These radio bursts can be considered a direct diagnosis of MHD shocks in the solar atmosphere. These bursts can be used to study the kinematics, energetics, and dynamics of the associated eruptive events very close to the Sun \citep{mancuso2004coronal, Anshu2017a, Morosan2019a, kandekar2025limitations}. 
 %Ramesh2010a
 Radio sources can provide important insights into the underlying physical mechanisms that drive the small and large-scale eruptions on the Sun \citep{gopalswamy2013height, 2014ApJ...787...59C, kumari2019direct, 2018ApJ...868...79C, Ramesh2023a, 2025arXiv250512932Z}. 
 %Ma2020
 Several studies have shown that the radio source shapes, sizes and evolution can be used to probe plasma turbulence in the solar corona \citep{mugundhan2017solar, kontar2017}. 

However, the smallest radio sources that can be observed in the solar corona have been less explored due to several observational challenges \citep{Bastian1994, Thejappa2008}. At metric wavelengths, it is practically difficult to obtain arcsecond spatial resolutions (comparable to visible wavelengths) due to the need of larger baselines of radio interferometers \citep{Monnier2013} and irregular refraction of radio waves in the solar corona due to turbulence \citep{2020ApJ...898...94K}. There have been a few occultation observations during solar eclipses at frequencies below 200 MHz, where the results indicate that sources sizes of $\sim1^\prime$ can be observed in the solar atmosphere \citep{Letfus1967, Kathiravan2011}. Hence, the largest baselines required for imaging the active Sun at frequencies $\le 200$ MHz needs to be investigated \citep{Aubier1971, Schmahl1994, Bastian2004} to obtain concrete information about the size of radio sources observed in the solar corona. 

With the recent advancement in the low frequency radio instruments (both solar and non-solar dedicated) such as the Mingantu Spectral Radio Heliograph \citep[MUSER;][]{Yan2021}, the LOw Frequency ARray \citep[LOFAR;][]{Haarlem2013}, the Murchison Widefield Array \citep[MWA;][]{Tingay2013}, and the upgraded Giant Metrewave Radio Telescope \citep[GMRT;][]{Gupta2017}, it is possible to observe the Sun in low frequencies with arcsecond angular resolutions. There have been a few studies in the past to determine the source sizes of radio bursts with these radio interferometers. \cite{Zhang2020a} studied the fundamental and harmonic emissions from a type III burst using interferometric observations with LOFAR and found that the source sizes were $\sim 25^\prime$ at $26.5$ MHz. Recently, \cite{Dabrowski2023} studied type IIIb and U solar bursts with LOFAR as well. Previously, 
\cite{Mercier2006} combined two radio interferometers in (GMRT in India and the Nan{\c c}ay Radioheliograph \citep[NRH;][]{Kerdraon1997} in France) and obtained the source size of $\sim 1^\prime$ at 326 MHz for solar noise storms.
%\cite{Murphy2021} studied type III bursts with LOFAR interferometric observations and found that the sources were $\sim 18.8^\prime$ at $36.7$ MHz. 
\citep{Gordovskyy2022} studied nine radio bursts (including two type II bursts) with LOFAR's tied-array beam mode. They used empirical methods to remove instrumental and ionospheric effects on the source sizes and found that the sizes can be $5^\prime-30^\prime$ between $45-30$ MHz. 
The smallest source size that has been reported at low frequencies by interferomteric observations are $\sim 15\arcsec$ reported by \cite{Mugundhan2018} at 55 MHz.
%and \cite{Cairns2018} at 107 MHz. 
Several other authors have also reported the source sizes to be $\sim 15-30\arcmin$ between 100-10 MHz \citep{Zhang2020a, Murphy2021}. However, almost all these studies are focused on solar noise storms or type III radio bursts. A recent interferometric study showed that increasing the baselines of radio observations and thus, increasing the spatial resolution, reveals new details in radio images of type II bursts, where an elongated single radio source in low-resolution images is in fact composed of two separate neighbouring  sources in high-resolution images \citep{morosan2025}. Thus, it is important to quantify the size of radio sources in the case of type II bursts in the solar corona. 

In this article, we take advantage of LOFAR's combined core (24 stations) and remote (14 stations) baselines, with a capability of providing a maximum baseline of $\sim 84~km$, to study type II radio source sizes and their shapes. For this, we used the LOFAR observations for three type II bursts: i) Aug 25, 2014  %at starting at 15:03 UT;
ii) Oct 16, 2015 (A)
%starting at 12:50 UT; 
and iii) Oct 16, 2015  (B).
%starting at 13:25 UT. 
This article is organised as follows: Section \ref{sec:section2} describes the LOFAR interferometric observations of these events; Section \ref{sec:section3} contains the data analysis methods and results; and in Section \ref{sec:section4}, we discuss the results and conclude the article. 
%great spectral, temporal and spatial resolutions.

\begin{table*}[!ht]
\centering
\caption{Details of the three type II radio bursts interferometric observations with LOFAR}
\label{tab:table1}
\begin{tabular}{|c|c|cc|cc|cc|cc|}
\hline
\multirow{2}{*}{Date} & \multirow{2}{*}{Details} & \multicolumn{2}{c|}{Frequency (MHz)} & \multicolumn{2}{c|}{Time (UT)}     & \multicolumn{2}{c|}{Imaging}   & \multicolumn{2}{c|}{Feature} \\ \cline{3-10} 
                      &                          & \multicolumn{1}{c|}{Start}   & End   & \multicolumn{1}{c|}{Start} & End   & \multicolumn{1}{c|}{LBA} & HBA & \multicolumn{1}{c|}{F}  & H  \\ \hline
25/08/2024            & F-H                      & \multicolumn{1}{c|}{170}     & 20    & 

\multicolumn{1}{c|}{15:01} & 15:31 & \multicolumn{1}{c|}{N}   & Y   & \multicolumn{1}{c|}{N}  & Y  \\ \hline
16/10/2015 (A)        & F-H                      & \multicolumn{1}{c|}{178}     & 30    & \multicolumn{1}{c|}{12:51} & 13:01 & \multicolumn{1}{c|}{Y}   & N   & \multicolumn{1}{c|}{Y}  & Y  \\ \hline
16/10/2015 (B)        & F-H                      & \multicolumn{1}{c|}{169}     & 33    & \multicolumn{1}{c|}{13:25} & 13:34 & \multicolumn{1}{c|}{Y}   & N   & \multicolumn{1}{c|}{Y}  & Y  \\ \hline
\end{tabular}
\end{table*}

\section{Data and Observation}
\label{sec:section2}

\subsection{Event 1: August 25, 2014  }
\begin{figure}[t]
\resizebox{0.5\hsize}{!}
{\includegraphics[width=\hsize]{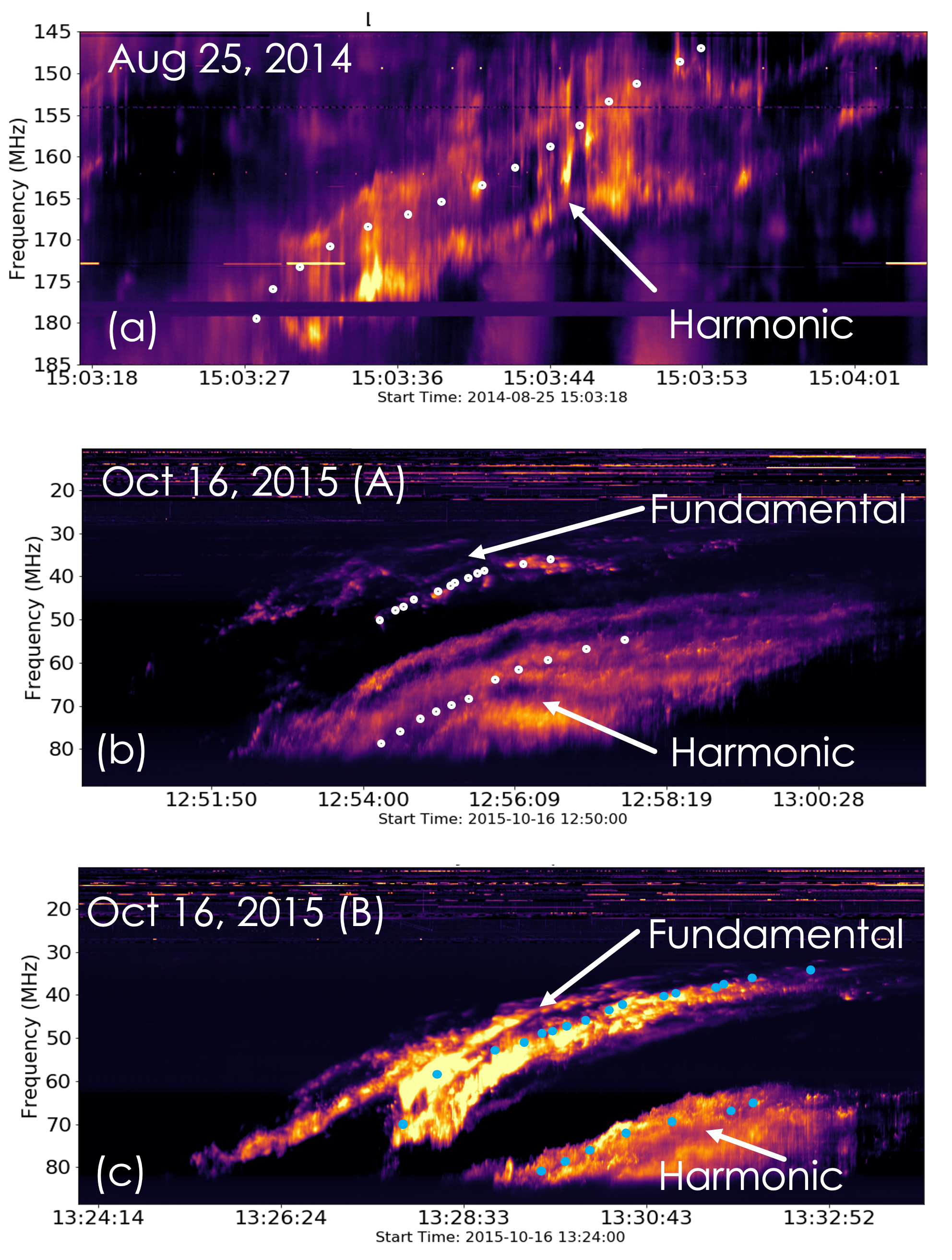}}
\caption{The dynamic spectra of: (a) Harmonic band of the type II burst observed with HBA of LOFAR on Aug 25, 2014; (b) Fundamental-Harmonic of the type II bursts observed with LBA of LOFAR on Oct 16, 2015 (A). The upper part of the spectra (HBA) is not shown here; (c) Same as (b) but for Oct 16, 2015 (B) event. The arrow shows the fundamental-harmonic bands of these type II bursts. The 'white' and 'cyan' marks are the time-frequency where interferomteric radio imaging were done and are shown in subsequent figures in this article. The imaging time and frequency were chosen in such a way to cover both fundamental and harmonic bands of type II bursts (if available).}
 \label{fig:figure1}
\end{figure}

A fundamental-harmonic (F-H) type II burst was observed with LOFAR on 25 Aug 2024 in both bands, high band antennae (HBA; 110–240 MHz), and low band antennae (LBA; 10–90 MHz) \citep{Haarlem2013}. The start and end frequencies of this type II burst were $\approx$ 170 MHz ($\approx$ 15:01 UT) and $\approx$ 20 MHz ($\approx$ 15:31 UT), respectively. Though LOFAR operates in spectral mode for both LBA and HBA, at present, the interferometric observations are only in one of the band. For this type II burst, the interferometric imaging observations were taken with the HBA, which corresponds to the harmonic band. The dynamic spectra for the HBA band is shown in Figure \ref{fig:figure1}(a). 

\subsection{Event 2: October 16, 2015 (A)}
Another F-H type II burst was observed with LOFAR both bands, HBA and LBA on Oct 16, 2015. The start and end frequencies of this type II burst were $\approx$ 178 MHz ($\approx$ 12:51 UT) and $\approx$ 30 MHz ($\approx$ 13:01 UT), respectively. For this type II burst, the interferometric imaging observations were taken with the LBA, which had both F-H bands. The dynamic spectra for the LBA band is shown in figure \ref{fig:figure1}(b). There was another type II burst on the same day, hence we will be referring to this event as 2015 October 16(A) here onwards in this article. 

\subsection{Event 3: October 16, 2015 (B)}
The third type II burst was observed on Oct 16, 2015 with both HBA and LBA bands of LOFAR. This was also a F-H pair. The start and end frequencies of this type II burst were $\approx$ 169 MHz ($\approx$ 13:25 UT) and $\approx$ 33 MHz ($\approx$ 13:25 UT), respectively. For this type II burst, the interferometric imaging observations were taken with the LBA, which had both F-H bands. The dynamic spectra for the LBA band is shown in figure \ref{fig:figure1}(c). We will be referring to this event as 2015 October 16(B) here onwards in this article. 

For Aug 25, 2014, the interferometric observations were obtained with 24 core stations and 12 remote stations, which provided in total {595} cross-correlation baselines.  Two sub-array-pointing beams were configured to observe the Sun (target) and Taurus-A (calibrator). For Oct 16, 2015, the interferometric observations were obtained with 23 core stations and 12 remote stations, which provided in total 561 cross-correlation baselines. Two sub-array pointings were configured to observe the Sun (target) and Virgo-A (calibrator). The temporal and spectral resolution of these interferometric observations were 0.25 s and 195.3 kHz, respectively. This high resolution unique LOFAR interferometric observation provided high-time and frequency resolution images of both F-H bands in the type II burst as well as the possibility to estimate the radio source sizes across the observed frequency ranges. The first event was studied in more detail by \cite{Jasmina2020} where the authors studied the fine structure of this type II. The second event was studied from shock origin point by \cite{Maguire2021} where the authors find that the type II originated from a jet driven piston shock. In the present study, we are reporting the type II source sizes and shapes for these events. 
%Figure \ref{fig:figure1} shows the dynamic spectra of the three type II bursts observed with LOFAR interferometric mode observations in the frequency range where the imaging observations were available. 

\section{Data Analysis and Results}
\label{sec:section3}
We determined the locations and source sizes of the type II bursts (both F-H) if available)  using the LOFAR interferometric observations. The dots marked in Figure \ref{fig:figure1} are the time-frequency points for which we performed radio imaging during these three events. For Event 1 (August 25, 2014) and Event 2 and 3 (October 16, 2015), the calibrator was Turaus A and Virgo A, respectively. We selected radio frequency interference (RFI) free bands for imaging. We used the Default Pre-Processing Pipeline (DPPP) for interferometric imaging \citep{vanDiepen2018}. Table \ref{tab:table1} shows the LOFAR bands and type II sources that were imaged from these datasets. 

LOFAR provides the raw visibility data in the form of Measurement Sets (MS) \citep{Haarlem2013}. MS is structured as a directory containing tables and sub-directories storing both data and metadata \citep{2007ASPC..376..127M}. The primary table, includes information such as the (u,v,w) coordinates in its UVW column, which are essential for correlating data from different antennas. Initially, we conducted manual inspection of the raw visibilities to identify and flag bad channels, antennas, and baselines. Following this, calibration procedures were applied using the observed calibrators specific to each event (see previous paragraph for details). The pre-processing and calibration as using the tool Default Pre-Processing Pipeline (DPPPP) as described in  \cite{2018ascl.soft04003V}. These calibration steps involved correcting for amplitude, phase, and bandpass variations across the observed frequencies. Once calibrated, the WSCLEAN algorithm  was employed for further data processing \citep{2014MNRAS.444..606O}. WSCLEAN performs image-domain de-convolution, utilizing the CLEAN algorithm, and subsequently restores the images from the calibrated visibility data. This iterative process was used to enhance the signal-to-noise ratio (SNR) while preserving the spatial resolution necessary for detailed imaging of the radio burst.

\begin{figure}
\resizebox{0.5\hsize}{!}
{\includegraphics[width=\hsize]{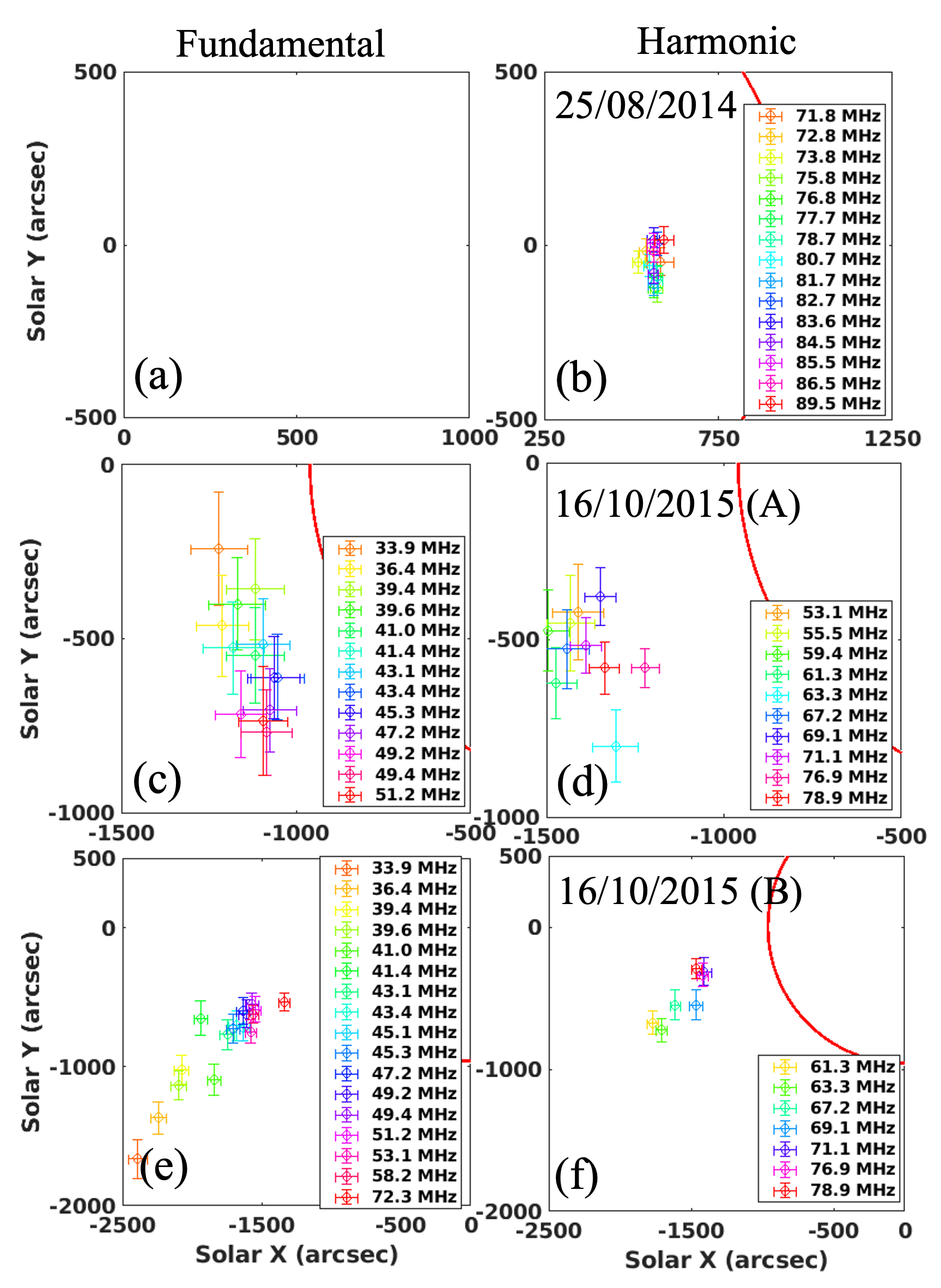}}
\caption{The centroid location of the type II radio burst sources for all the three events studied in the present work (row-wise). The fundamental and harmonic radio sources are plotted column-wise. The extent of these sources in the major and minor directions is represented by the bars centered on each centroid in the x and y directions, respectively.  The `red' circles show the visible Sun. Since there was no interferomteric observations for the fundamental band of the first type II, we have images for the harmonic band only. These images correspond to the 'white' and 'cyan' markings on the three spectra shown in figure \ref{fig:figure1}. }
\label{Fig:figure2}
\end{figure}

\begin{figure}
\resizebox{0.5\hsize}{!}
            {\includegraphics[width=0.8\textwidth,clip=]{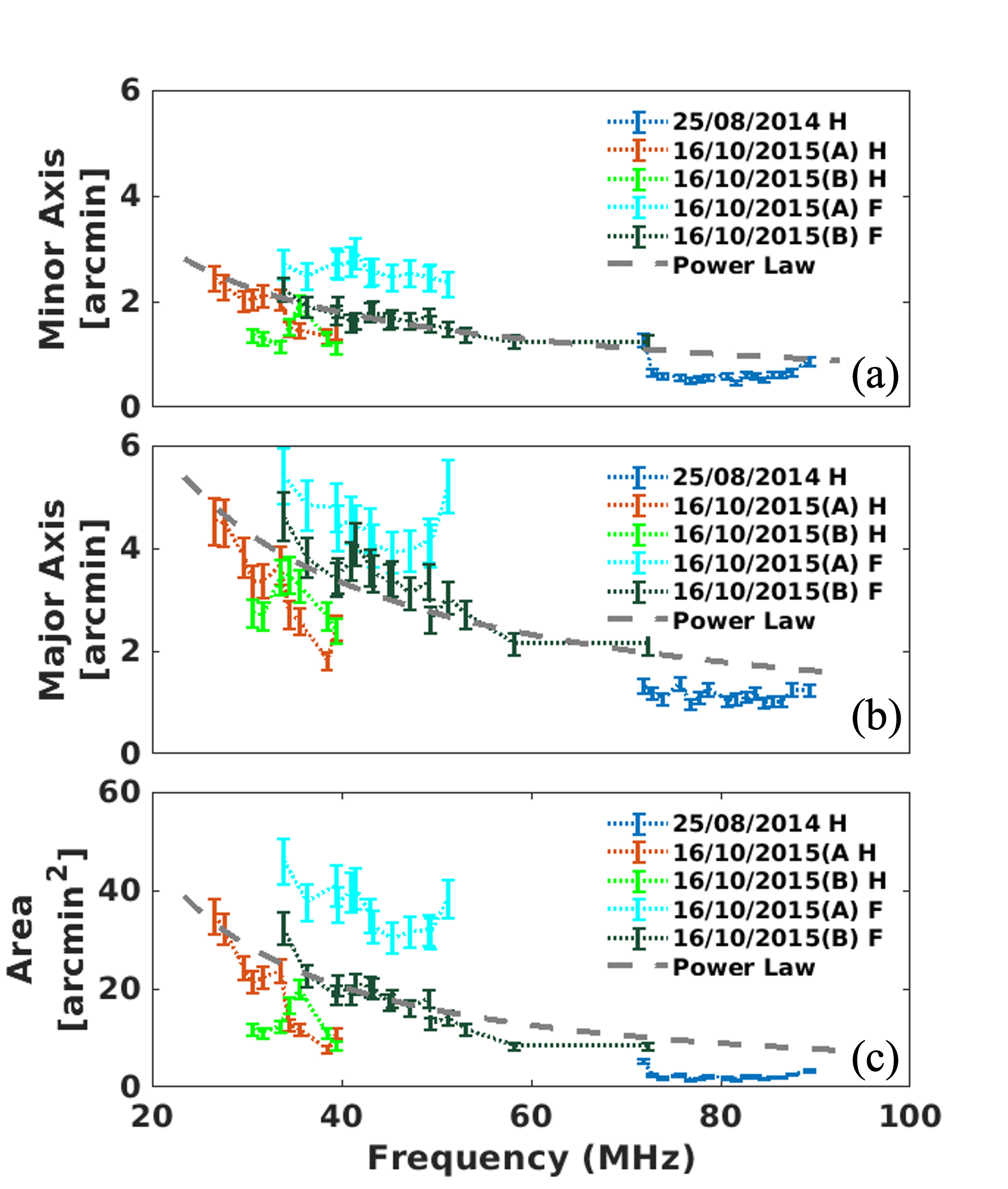}}
      \caption{The sizes (major and minor extent [1/e]) and area  of the radio sources for all the three type II bursts studied in the present work: (a) radio sources extend in equatorial direction; (b) radio sources extend in polar direction; (c) area of the radio sources. Different events and data corresponding to their features are shown in different colors. The 'grey' dashed line is a combined power-law fit to the data. The errorbars denote the uncertainty in source sizes following the Gaussian fit to the radio sources to estimate their sizes. }
         \label{Fig:figure3}
   \end{figure}

Figure \ref{Fig:figure2} shows the location of the type II bursts from this study. We estimated the source sizes and its positions using a 2D Gaussian fit for calibrated and corrected solar images. The source sizes were calculated using 2D Gaussian fits, and then errors were estimated again using the maximum brightness temperature, noise floor and beam sizes at different frequencies (see for example the methods used in \citealt{Morosan2019a}). The x and y extent bars in Fig.~\ref{Fig:figure2} denote the size of the source in the major and minor axes directions, respectively, while the error bars in \ref{Fig:figure3} denote the uncertainty in source sizes following the Gaussian fit. 

We fitted a power law equation to estimate the frequency dependency of the source sizes on the solar atmosphere. The power law exponent for the major axis varied, with values of $\approx1.5$ for harmonic and $\approx0.75$ for fundamental sources. In Figure \ref{Fig:figure3}, we show the source sizes estimated for all three events analyzed in this study, along with their corresponding power law exponents for both minor and major axes. Our results indicate that for fundamental lanes associated with type II bursts, estimated source sizes range between $1-5'$ for frequency range $30-70~MHz$. In contrast, for harmonic sources, sizes varied around $\approx1'$ and $\approx 2'$ at higher ($180-145~MHz$) and lower LOFAR bands ($85-55~MHz$), respectively. These sizes slightly exceed the beam sizes observed at these frequencies during interferometric observations with LOFAR using the antenna set of core plus remote array with the longest baseline of {$\approx13~km$}.

For the minor axis, fundamental source sizes ranged from $\approx 2-5'$, while harmonic sources were approximately $\approx 2'$ in size. Notably, the source sizes in both major and minor axes directions were significantly bigger than the beam size in the East-West direction during interferometric observations, suggesting limitations imposed by baseline distance or snapshot imaging where not enough signal would have been detected by the farther remote stations in the limited time-frame. \cite{mugundhan2018spectropolarimetric} had reported source sizes be as low as $\approx 15''$ at 53 MHz, compared to this, we got almost $\approx 8\times$ bigger sizes and one of the reasons for that are smaller baselines. Another reason for out result could be also high level of the solar activity at the time of studied events, during which we expect scattering effects to be more pronounced. While comparing these source sizes obtained with interferometric observations to the scattering removed tied-array beam observation by \cite{Gordovskyy2022}, we report significantly lower source sizes for type II bursts observed with LOFAR's interferometric mode.

\section{Discussion and Summary}
\label{sec:section4}
The smallest angular dimensions of compact radio sources in the solar atmosphere that can be observed is still an open question. As it has been evident with large baseline radio instruments, many small sources can be observed. The radio source sizes play a crucial role in understanding the dynamics of density turbulence within the solar corona, specially at low frequencies and its effect on the angular extent of radio sources. The extent of the interferometric baselines necessary to observe radio sources in the solar atmosphere must be sufficiently large to put a constrain on the observable smallest compact radio source. The 2D geometry of low frequency radio sources and their frequency dependence has been under-explored due to the limited spatial resolution of radio imaging telescopes. 
%The intricate geometrical properties of low-frequency solar radio sources and their variation with frequencies remain a subject of limited studies, largely attributable to the inherent constraints of spatial resolution in solar observations.

Here, we presented a comprehensive study of multiple type II radio bursts across various frequency bands, to understand their source characteristics. We analysed source shapes by approximating the derived intensity distributions through the application of 2D Gaussian profiles using elliptical 1/e contours. The three events shown here had fundamental and/or harmonic imaging observations available with LOFAR. In our studies we looked into the shape and sizes of these radio bursts. The sizes of the radio sources were estimated by the extent of the source, which depends upon several factors, including the peak brightness, root mean square (rms) noise floor, and beam size at each frequency. 
We note that source sizes can also vary on sub-second ($\approx 100$ ms) scales. In the current study, we focused on the variation of source characteristics with frequency across different events rather than on temporal variations at sub-second scales.
%Nevertheless, notable differences emerge in the positions of harmonic frequencies, which appear to be situated at higher elevations. This discrepancy suggests that higher frequencies may be subject to less significant effects of scattering (cite someone??).

Figure \ref{Fig:figure2} indicates that in a case of event 3, and clearly radially-outward propagating radio sources, i.e. approximately along the radially stretching magnetic field lines (see panels (e) and (f)) the obtained radio source sizes are rather small and compact. In a case of event 2, the positions of the radio sources at subsequent frequencies indicate shock wave propagation towards the observer under angle of $< 90^o$ (see panels (c) and (d)) indicating that the large spread of radio source sizes could be resulting from the radio emission propagating in the region with the complex configuration of the magnetic field. On the other hand, very compact and almost completely superposed radio source positions in a case of Event 1, indicate type II bursts propagation fully along the Sun-Earth line. However, we note that without having a 3D radio source positions we can only conclude that the obtained trajectories and the radio source sizes for studied Events indicate both different scattering effects and possibly different orientation of the underlying coronal magnetic field.

\begin{figure}
 \resizebox{0.5\hsize}{!}
 {\includegraphics[width=0.99\textwidth,clip=]{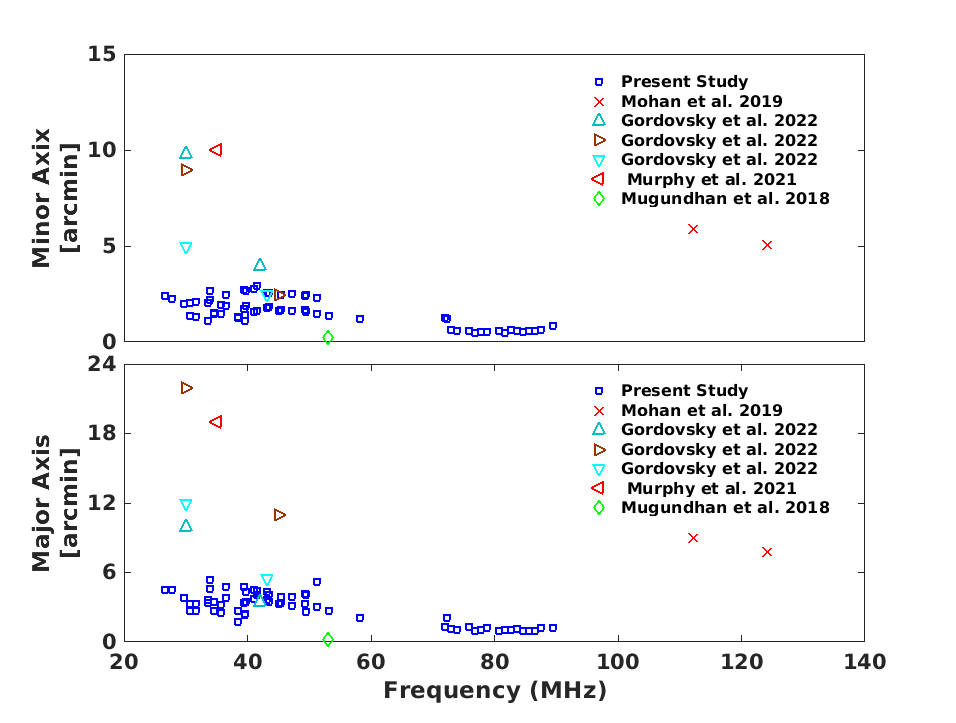}}
 \caption{Combined plot of source sizes with various previous studies. 'Blue' squares are the source sizes for fundamental-harmonic lanes of three type II radio bursts studied in the present case. Notably, the source sizes (both minor and major) obtained in the present study is smaller than all the previous studies with different instruments and different mode of observations (except \cite{Mugundhan2018}).}
  \label{Fig:figure4}
\end{figure}

Figure \ref{Fig:figure4} shows a combined plot of some of the previous studies when compared to the results obtained with the present study. We compared our results with the type II bursts observed with LOFAR tied-array beam observations of type II radio bursts \citep{Gordovskyy2022}, type III bursts observed with MWA \citep{Mohan2019}, type III radio bursts observations with interferometric mode of LOFAR \citep{Murphy2021} and very long baseline interferomtery (VLBI) of noise storm \citep{Mugundhan2018}. Notably, the source sizes obtained in the present study is smaller than previous studies except that of \cite{Mugundhan2018}, where the baseline was $\approx 200$~km. 

When comparing our findings with those of \cite{Gordovskyy2022}, which is the only other study available where type II sources sizes were analyzed, a disparity was noticed: the observed sources in the present study exhibit significantly smaller dimensions, without the application of any scattering removal techniques. Despite this discrepancy, both our results and those of \cite{Gordovskyy2022} align with models of anisotropic radio-wave scattering within the solar corona, when using different scattering parameters to determine the level of turbulence \citep[e.g.,][]{kontar2023}. In comparison with other LOFAR observations, the source sizes in the present study have smaller areas than those reported by \citet{Dabrowski2023} for type III radio bursts, that also used LOFAR's interferometric mode in a similar configuration. Further studies are thus necessary to quantify if scattering effects are generally larger for type III bursts compared to type II  bursts or if these effects change significantly with solar activity.
 %This suggests that tied-array beam observations may be limited by the resolution of LOFAR's core stations which are capapble of producing these beams, where the useable baseline is only 2~km.  

%Previously, \cite{mugundhan2018spectropolarimetric} has demonstrated markedly reduced source sizes for type III radio bursts (as small as $\approx 15''$) during periods of solar cycle minima. Furthermore, \cite{mugundhan2018spectropolarimetric} and concluded that radio sources are less impacted by scattering phenomena during solar minima.

Previously, \cite{mugundhan2018spectropolarimetric} has demonstrated that the type III radio bursts source sizes, during periods of solar cycle minima,  can be as small as $\approx 15''$.  The authors connected this effect to the smaller impact of the scattering processes on the radio source sizes at the time of the low level of solar activity when the radio emission propagates through less structured corona. We note that LOFAR observations in the present study were taken during the solar maximum phase. It is important to note that radio scattering is inherently anisotropic, which can influence observed source sizes \citep{2023ApJ...956..112K}. This anisotropy, alongside other scattering effects, may contribute to the broadening of solar radio sources and should be considered when interpreting size measurements. Despite this, our present findings indicate notably smaller radio source sizes than previously reported.
%\iffalse
This could imply three different scenarios:
\begin{enumerate}
    \item Scattering effects are generally smaller than previously expected.
    \item As the source would be embedded in the scattering medium itself, scattering could happen much closer to the vicinity of the source itself, resulting in the spherical waves from source being scattered, rather than plane waves, which could result in smaller angular sizes \citep{subramanian2011}.
    \item The level of the scattering experienced by the radio emission is depending not only on the density inhomogeneities of the ambient corona \citep{2023A&A...670A..20J} but  also on the orientation of the underlying magnetic filed.
\end{enumerate} 
%\fi
%%At present, due to observational constraints, neither of the aforementioned scenarios can be verified.
Figure \ref{Fig:figure3}c shows the area of the radio sources for the type II bursts studied in the present study matches closely with those from simulations by \cite{2025ApJ...978...73C} that found $\approx 29.7~{arcmin}^2$ in non-radial magnetic fields.
%reflecting anisotropic scattering in non-radial magnetic fields rather than reduced scattering effects. 
We note that similar source sizes were also reported in a different study for herringbones found in a type II burst \citep{2024A&A...683A.123Z}. %for herringbones type II radio bursts.}
%This underscores the robustness and efficacy of interferometric observations in determining the extent of the impact of scattering phenomena, even during periods of high solar activity.
Future high-resolution interferometric observations, during different stages of solar activity, have the potential to further constrain these intrinsically complex problem.

\begin{acknowledgements}
{A.K., D.E.M. and P.Z. acknowledge the University of Helsinki Three Year Grant. A.K is supported by an appointment to the NASA Postdoctoral Program at the the NASA Goddard Space Flight Center (GSFC). AK acknowledges the ANRF Prime Minister Early Career Research Grant (PM ECRG) program. D.E.M acknowledges the Academy of Finland project `RadioCME' (grant number 333859) and Academy of Finland project `SolShocks' (grant number 354409). E.K.J.K. and A.K. acknowledge the European Research Council (ERC) under the European Union's Horizon 2020 Research and Innovation Programme Project SolMAG 724391. E.K.J.K acknowledges the Academy of Finland Project SMASH 310445. All authors acknowledge the Finnish Centre of Excellence in Research of Sustainable Space (Academy of Finland grant number 312390). The authors wish to acknowledge CSC – IT Center for Science, Finland, for computational resources. This paper is based on data obtained with LOFAR \citep{Haarlem2013}, which is the Low Frequency Array designed and constructed by ASTRON. This research used version 4.1 \citep{sunpy_community2020} of the SunPy open source software package and version R2018b \citep{MATLAB} of MATLAB.}
\end{acknowledgements}

\begin{appendix}
\section{Raw images and visibilities}\label{sec:app}

To ensure potential over-cleaning, we took several precautions to ensure the quality of our images:

\begin{itemize}
    \item \textbf{Thresholding:} We carefully set the cleaning threshold to maintain the maximum brightness in the corrected data above the noise floor. This approach helps prevent over-cleaning while preserving faint signal details, consistent with standard practices in radio astronomy imaging tools such as AIPS, CASA, and WSClean.
    
    \item \textbf{Signal-to-Noise Ratio (SNR) Control:} We systematically monitored the SNR throughout the cleaning process to avoid cleaning at or below the noise level, which would risk over-cleaning the data.
    
    \item \textbf{Cleaning Algorithm:} We employed the CLEAN algorithm originally developed by Högbom (1974), specifically using the WSClean implementation by Offringa et al. (2014). This method has been widely adopted for over five decades in radio astronomy for its effectiveness in reconstructing complex source structures. Its reliability and robustness are well established within the community.
\end{itemize}

Recent studies \citep{morosan2025} have shown that increasing the baseline lengths in solar radio imaging of type II bursts results in higher image resolution and decreased source sizes, which remain larger than the WSClean restoring beam. Below we provide a sample calibrated image in the RA--Dec coordinate system. The peak brightness reaches approximately $10^9$ K, while the residual noise level is around $10^5$ K, resulting in a signal-to-noise ratio (SNR) of about 40 dB.

\begin{figure}
    \centering
    \includegraphics[width=0.5\linewidth]{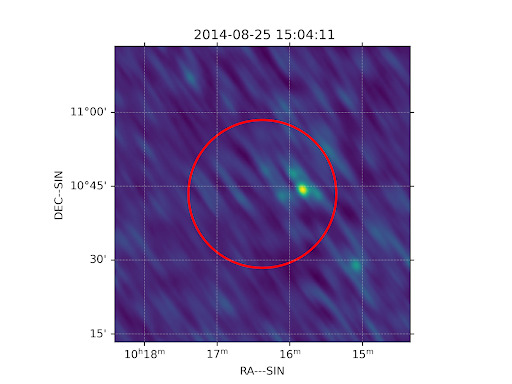}
    \caption{A raw radio image (corrected for phase, amplitude, and bandpass effects, but not cleaned) shown in the RA--Dec coordinate system. The red circle represents the solar photosphere.}
    \label{fig:method_nobeyama}
\end{figure}

In addition to image-domain analysis, we also examined the raw visibility data to ensure that the source sizes inferred from the interferometric images are consistent with the UV-distance behavior, both during quiet and active Sun conditions.

In these visibility plots, we observe that the quiet Sun visibility amplitudes drop close to the noise floor at UV distances of approximately 5~km. In contrast, for the radio bursts associated with the type II event, the visibility amplitudes reach the noise floor only at baselines of approximately 13~km. This clearly indicates that compact structures with sizes as small as $\sim0.5''$ are present at the higher frequencies during the type II emission, consistent with the sizes inferred from the interferometric images presented in this study.

Moreover, since our analysis involves a large number of images generated through a semi-automated imaging pipeline, it is more straightforward and efficient to handle the cleaned image products rather than examining the visibility amplitudes for each time and frequency point individually.

\begin{figure}
    \centering
    \includegraphics[width=0.45\textwidth]{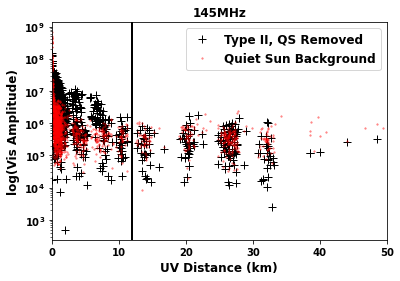}
    \caption{Visibility amplitudes versus UV distance for type II bursts (black) and quiet Sun images (red). Here, the quiet Sun is averaged over 2 minutes to enhance the contrast. After the black curve flattens, the excess flux arises from increased noise due to the shorter (20-second) integration during the burst.}
    \label{fig:uvamp_diffintegration}
\end{figure}

\iffalse
\begin{figure}
    \centering
    \includegraphics[width=0.45\textwidth]{A5.png}
    \caption{Visibility amplitudes versus UV distance for type II bursts (red) and quiet Sun images (black) at different frequencies.}
    \label{fig:uvamp_differentfreq}
\end{figure}
\fi

\begin{figure}
    \centering
    \includegraphics[width=0.45\textwidth]{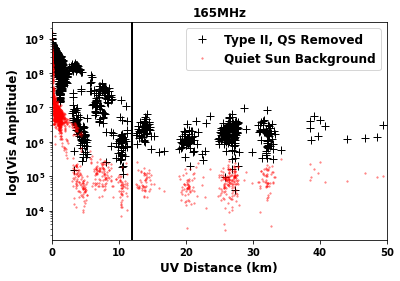}
    \caption{Same as above but at 165 MHz.}
    \label{fig:uvamp_differentfreq}
\end{figure}

\end{appendix}

\end{document}